# Nature of "Spontaneous Curvature" in Suspended Graphene

Yangfan Hu[1,*]

[1]Sino-French Institute of Nuclear Engineering and Technology, Sun Yat-Sen University, 510275 GZ, China

The nature of its intrinsic ripples[1, 2, 3] is the key factor for understanding the stability of suspended graphene, and for unraveling the long-standing theoretical debate of the existence of low-dimensional crystalline state[4, 5, 6]. The rippling morphology of graphene, discovered also in other two-dimensional (2D) materials[7, 8, 9, 10], has a profound impact on its electronic[11, 12, 13], mechanical[14, 15, 16] and chemical[9, 17, 18] properties. Actually, long before the discovery of graphene, rippling phenomena are widely observed in many different systems: for example, the roughing transition of crystalline interface[19], the "rippled" phase in biomembrane[20, 21] and crumpling of flexible sheet polymers modeled by "tethered surfaces"[22, 23]. The fascinating truth that ripples exist in so many membrane-like materials with significant difference implies possible existence of a universal physical mechanism which was unclear. We clarify that the intrinsic ripples in suspended graphene is composed of two parts owing to different formation mechanism. The first part is characterized by the "*spontaneous curvature*", which derives from the need to stabilize the soft ZA modes near long wavelength limit[24]. The second part is characterized by the "*thermal curvature*", which is caused directly by height fluctuation. By choosing the spontaneous curvature as the order parameter of the system, we establish the Landau theory modfied by thermal fluctuation for "*wrinkling transition*" of large sized graphene. **We find that as temperature rises from 0K, a second order phase transition occurs at a size dependent critical temperature $T_c$, which corresponds to a change of equilibrium configuration from a flat state to a rippling state. Interestingly, the order parameter is stablized as temperature increases,** and the phase transition is associated with a sudden increase of equilibrium bond length as well as a vanishing "intrinsic bending rigidity". The results obtained suggest

*Corresponding author. E-mail: huyf3@mail.sysu.edu.cn

**that the interplay between the rippling morphology and the elementary excitations is vital for understanding the behavior of any suspended 2D materials. The concepts and theory developed here is of general significance at least for tethered membranes.**

Due to the complex microscopic structure and constituents of flexible membranes, theoretical concerns about their intrinsic ripples are usually developed upon continuum medium hypothesis[6, 23, 25] without including any microscopic feature. For the same reason, a lot of effort has been made on analyzing the effect of geometrical deformation and defects (e.g. the KTHNY theory[26, 27, 28, 29] for 2D melting) on ripple formation, while the interplay between ripples and the elementary excitations is not considered. In this case, solving the ripple problem in graphene is particularly valuable since its atomic interaction and vibrational behavior can both be explicitly obtained[15, 30, 31], which permits discussion on the applicability of existing membrane theories[3, 32]. Through phonon analysis of graphene[24], it is already known that the ZA modes near long wavelength limit are softened when the equilibrium lattice constant is shortened due to negative thermal expansion (NTE). This means large sized ripples spontaneously appear with shape determined by the soft modes[33, 34]. These "*spontaneous ripples*" refer to change of the static equilibrium configuration and should be distinguished from the "*thermal ripples*" due to height fluctuation. While too much attention is paid to the effect of thermal ripples, we believe that the spontaneous ripples have more fundamental influence on the morphology and properties of graphene.

To study how the soft modes are stabilized by presence of the spontaneous ripples, we need to solve the dispersion relation for rippling graphene at long wavelength limit. Since the spontaneous ripples derive from soft ZA modes near long wavelength limit, the wavelength of these soft modes is significantly larger than lattice spacing. In the absence of any mechanical load, the ripple size of suspended graphene should be isotropic in the surface considered. Therefore, the structure of rippling graphene can be regarded as being composed of many "local regions" with two equal principal curvatures denoted by $\kappa$. One should notice that the idea developed here resembles Kadanoff's block spin transformation for 2D Ising models [35], and that a similar idea had been mentioned before to describe a rippling graphene[36]. To characterize a local region as slightly perturbed flat plane, let us define a nondimensionalized curvature $\bar{\kappa}$ as $\bar{\kappa} = a\kappa$ where $a$ denotes the lattice constant of graphene. For different local regions, $\bar{\kappa}$ can take different values and different signs. Yet, due to the absence of macroscopic bending moment, the average of $\bar{\kappa}$ over the whole material should be zero and the sign of $\bar{\kappa}$ for neighboring regions of one local region with positive $\bar{\kappa}$ should be negative to ensure equilibrium of local bending moment. Define the *spontaneous curvature* as

$$\tilde{\kappa} = \frac{1}{S}\int_S |\bar{\kappa}|ds, \tag{1}$$

where $S$ denotes the total area occupied by the 2D material. At equilibrium $|\bar{\kappa}|$ of an arbitrary local region would fluctuate around $\tilde{\kappa}$. Consider the lattice dynamics for the acoustic branches at long wavelength limit, the dynamical matrix can be expanded as

$$\mathbf{D}^{acoustic}\begin{pmatrix}\mathbf{q}\\k,k'\end{pmatrix} = \mathbf{D}_0\begin{pmatrix}\mathbf{q}\\k,k'\end{pmatrix} + \mathbf{D}_2\begin{pmatrix}\mathbf{q}\\k,k'\end{pmatrix}\tilde{\kappa}^2, \tag{2}$$

where $\mathbf{q} = q[\cos\alpha, \sin\alpha]^T$ denotes the wave vector in the 2D reciprocal space, $\mathbf{D}_0$ denotes the dynamical matrix[37] of graphene in flat configuration and $\mathbf{D}_2$ represent the second order disturbance of the dynamical matrix due to $\tilde{\kappa}$ (for derivation see the Supplementary Information).

Through matrix perturbation theory[38], the expression of the acoustic branches phonon frequencies at long wavelength limit can be obtained by solving the eigenvalue of $\mathbf{D}^{acoustic}\begin{pmatrix}\mathbf{q}\\k,k'\end{pmatrix}$ as

$$\omega_{ZA}^2 = b_3^* q^2 + d_3 q^4, \quad b_3^* = b_3 + \overline{h_3}\tilde{\kappa}^2, \tag{3a}$$

$$\omega_{LA}^2 = b_1^* q^2, \quad b_1^* = b_1 + \overline{h_1}\tilde{\kappa}^2, \tag{3b}$$

$$\omega_{TA}^2 = b_2^* q^2, \quad b_2^* = b_2 + \overline{h_2}\tilde{\kappa}^2, \tag{3c}$$

where $b_i$ and $\overline{h_i}, (i = 1,2,3)$ can be expressed in terms of the force constants. The expressions and details of the solution process can be found in Supplementary

Information. According to eqs. (3a-c), when spontaneous curvature appears, coefficients of the second order term of $q$ in the dispersion relations for all the acoustic branches at long wavelength limit are renormalized. When describing the carbon-carbon potential with the empirical bond order potential LCBOPII[30], we have after evaluation $\overline{h_1} < 0, \overline{h_2} < 0$ and $\overline{h_3} > 0$. For flat graphene, soft modes appear when $b_3 < 0$, i.e., when the equilibrium lattice constant $a$ becomes smaller than the static value $a_0$ (here $a_0$ is determined by $\frac{\partial \Phi_i}{\partial a} = 0$, where $\Phi_i$ denotes the potential energy possessed by the $i$th atom)[24]. Eq. (3a) provides solid evidence that the soft modes can be stabilized by spontaneous curvature.

At given temperature and zero mechanical loads, the equilibrium lattice constant and spontaneous curvature take values that minimizes the Helmholtz free energy. Since it's well known that thermal fluctuation[32, 39] plays a significant role in ripple formation of any 2D materials, here the Helmholtz free energy per atom is described within the self-consistent harmonic approximation (SCH)[40, 41, 42] as

$$f_{SCH} = \langle \Phi_i \rangle + \frac{k_B T}{N} \sum_{\mathbf{q},\lambda} \ln(1 - e^{-\frac{\hbar \omega_{\mathbf{q}\lambda}}{k_B T}}) - \frac{1}{2N} \sum_{\mathbf{q},\lambda} \hbar \omega_{\mathbf{q}\lambda} \left( e^{\frac{\hbar \omega_{\mathbf{q}\lambda}}{k_B T}} - 1 \right)^{-1}, \quad (4)$$

where $N$ denotes the number of atoms in the system and $\langle ... \rangle$ denotes statistical average over the states described by the Hamiltonian of the system under harmonic interactions. One should notice that in eq. (4), it is assumed that $\langle \Phi_i \rangle$ describes the ground state Helmholtz free energy per atom (i.e. the free energy caused by zero point vibration is incorporated in $\langle \Phi_i \rangle$)[43]. If the contribution of optic phonon branches to the free energy is neglected, eq. (4) can be reformulated within the Debye model as

$$\begin{aligned} f_{SCH} = \langle \Phi_i \rangle &+ \frac{1}{N} \int_0^{\omega_{in}} \left[ k_B T \ln(1 - e^{-\frac{\hbar \omega}{k_B T}}) - \frac{\hbar \omega}{2} \left( e^{\frac{\hbar \omega}{k_B T}} - 1 \right)^{-1} \right] g_{in}(\omega) d\omega \\ &+ \frac{1}{N} \int_0^{\omega_{out}} \left[ k_B T \ln(1 - e^{-\frac{\hbar \omega}{k_B T}}) - \frac{\hbar \omega}{2} \left( e^{\frac{\hbar \omega}{k_B T}} - 1 \right)^{-1} \right] g_{ZA}(\omega) d\omega, \end{aligned} \quad (5)$$

where $g_{in}(\omega)$ and $g_{ZA}(\omega)$ denote respectively the density of state for in-plane acoustic branches and ZA modes using the dispersion relation at long wavelength limit. According to eqs. (3a-c), we have

$$g_{in}(\omega) = \frac{S}{2\pi}\left(\frac{1}{\langle b_1^*\rangle} + \frac{1}{\langle b_2^*\rangle}\right)\omega, \tag{6a}$$

$$g_{ZA}(\omega) = \frac{S}{2\pi}\frac{\omega}{\sqrt{\langle b_3^*\rangle^2 + 4\langle d_3\rangle\omega^2}}, \tag{6b}$$

where $S$ denotes the total area of the 2D material. In eq. (5) $\omega_{in} = 4\sqrt{\frac{\pi\langle b_1^*\rangle\langle b_2^*\rangle}{s_0(\langle b_1^*\rangle+\langle b_2^*\rangle)}}$ and $\omega_{out} = 2\sqrt{\frac{2\pi}{s_0}\left(\frac{8\langle d_3\rangle\pi}{s_0}+\langle b_3^*\rangle\right)}$ denote the two Debye frequencies for in-plane acoustic modes and for ZA modes, respectively. Here $s_0 = \frac{3\sqrt{3}}{2}a^2$ denotes the area occupied by one lattice cell. Through simple analysis of eqs. (5, 6b), it is found that to minimize the free energy the spontaneous curvature has a definite expression

$$\tilde{\kappa}^2 = \begin{cases} -\frac{\langle b_3\rangle}{\langle h_3\rangle}, & if\ \langle b_3\rangle < 0 \\ 0, & if\ \langle b_3\rangle \geq 0 \end{cases}. \tag{7}$$

The analysis is given as follow: Assume that for $\langle b_3\rangle < 0$, $\tilde{\kappa}^2 = -\frac{\langle b_3\rangle}{\langle h_3\rangle}$ is increased by $d\tilde{\kappa}^2 = \frac{\delta}{\langle h_3\rangle}$, where $\delta$ is positive and infinitely small, we have $dg_{ZA}(\omega) = \frac{S}{2\pi}\left(\frac{\omega}{\sqrt{\delta^2+4\langle d_3\rangle\omega^2}} - \frac{1}{2\sqrt{\langle d_3\rangle}}\right) < 0$ for any $\omega$. Moreover, for $\omega \to 0$, $dg_{ZA}(\omega) \to -\frac{S}{2\pi}\frac{1}{2\sqrt{\langle d_3\rangle}}$. On the other hand, $dg_{in}(\omega) = -\frac{S}{2\pi}\left(\frac{\langle h_1\rangle}{\langle b_1^*\rangle^2} + \frac{\langle h_2\rangle}{\langle b_2^*\rangle^2}\right)\frac{\delta}{\langle h_3\rangle}\omega > 0$ is linear in $\delta$. Hence $dg_{in}(\omega)$ in the low frequency range is infinitesimal compared with $dg_{ZA}(\omega)$, and so the vibrational part of the free energy is increased because of $\delta$. Meanwhile, the potential part of the free energy per atom can be expanded for small $\tilde{\kappa}$ as

$$\langle\Phi_i\rangle = \langle\Phi_{i0}\rangle + \frac{9}{16}\langle c_a\rangle\tilde{\kappa}^2 + \frac{9}{256}(3\langle c_{a2}\rangle - \langle c_a\rangle)\tilde{\kappa}^4, \tag{8}$$

where $\Phi_{i0}$ denotes the potential for flat graphene and higher order terms are omitted. Since $\langle c_a\rangle > 0$ and $3\langle c_{a2}\rangle - \langle c_a\rangle > 0$ for graphene, we have $d\langle\Phi_i\rangle > 0$ for any given $\delta$. Therefore $df_{SCH} > 0$ for given $\delta$, which renders eq. (7). Substituting eqs. (6, 7) into eq. (5), after manipulation we have

$$\begin{aligned}f_{SCH} = &\langle\Phi_{i0}\rangle + \frac{9}{16}\langle c_a\rangle\langle\tilde{\kappa}^2\rangle + \frac{9}{256}(3\langle c_{a2}\rangle - \langle c_a\rangle)\langle\tilde{\kappa}^2\rangle^2 \\ &+ \frac{s_0(k_BT)^3}{4\pi\hbar^2}\left(\frac{1}{\langle b_1^*\rangle} + \frac{1}{\langle b_2^*\rangle}\right)J_1 + \frac{s_0(k_BT)^2}{8\pi\hbar}\frac{1}{\sqrt{\langle d_3\rangle}}J_2,\end{aligned} \tag{9}$$

where $J_1 = \int_0^{\Theta_{in}/T} \left[\ln(1-e^{-\xi}) - \frac{\xi}{2}(e^\xi - 1)^{-1}\right]\xi d\xi$ and $J_2 = \int_0^{\Theta_{out}/T}\left[\ln(1-e^{-\xi}) - \frac{\xi}{2}(e^\xi-1)^{-1}\right]d\xi$ can both be considered as constants below room temperature. Here $\Theta_{in} = \frac{\hbar\omega_{in}}{k_B}$ and $\Theta_{out} = \frac{\hbar\omega_{out}}{k_B}$ denote the in-plane Debye temperature and out-of-plane Debye temperature. At ground state, $\Theta_{in} \approx 2600K$ and $\Theta_{ZA} \approx 1400K$. In self-consistent phonon theory, the terms in $\langle...\rangle$ are evaluated as functions of vibrational displacements following Gaussian distribution, where its variance is related to the fluctuation of the nearest neighbor bond length. At long wavelength limit using the Debye model we have after deduction

$$\Lambda_{Z1} = \frac{a^2 s_0 k_B T}{8\pi m \langle d_3 \rangle} \ln\left[\frac{1 - e^{-\frac{\Theta_{out}}{T}}}{\lim_{\omega \to 0}\left(1 - e^{-\frac{\hbar\omega}{k_B T}}\right)}\right] + \frac{\hbar a^2}{2m\sqrt{\langle d_3 \rangle}}, \tag{10a}$$

$$\Lambda_{L1} = \frac{\hbar a^2}{3m}\sqrt{\frac{\pi}{s_0 \langle b_1^* \rangle}} + \frac{a^2 s_0 (k_B T)^3}{8\pi m \hbar^2}\frac{J_3}{\langle b_1^* \rangle^2}, \tag{10b}$$

$$\Lambda_{T1} = \frac{\hbar a^2}{3m}\sqrt{\frac{\pi}{s_0 \langle b_2^* \rangle}} + \frac{a^2 s_0 (k_B T)^3}{8\pi m \hbar^2}\frac{J_3}{\langle b_2^* \rangle^2}, \tag{10c}$$

where $\Lambda_{Z1}$, $\Lambda_{L1}$ and $\Lambda_{T1}$ denote respectively the fluctuation of the nearest neighbor bond length due to ZA, LA and TA modes. And $J_3 = \int_0^{\Theta_{in}/T} \xi^2 \left(\coth\frac{\xi}{2} - 1\right) d\xi$ in eqs. (10b-c) is finite and can be regarded as constant below room temperature.

We observe from eq. (10a) that for infinite sized graphene, $\Lambda_{Z1} \to \infty$. It's been proved long ago[5] that for 2D materials the in-plane displacement correlation diverges in the form $\langle [\mathbf{u}_{in}(\mathbf{R}) - \mathbf{u}_{in}(\mathbf{R}')]^2 \rangle \sim \ln|\mathbf{R} - \mathbf{R}'|$ as $|\mathbf{R} - \mathbf{R}'| \to \infty$. The divergence obtained in eq. (10a) is stronger since it denotes correlation of the out-of-plane displacement of two neighboring atoms with $|\mathbf{R} - \mathbf{R}'| = a$. Physically, this means that for infinite sized graphene with flat configuration, before in-plane melting occurs it will first undergo severe out-of-plane undulation. We thus obtain an important result: graphene with infinite size is extremely unstable at flat configuration due to out-of-plane fluctuation.

For material sample with finite size $L^2$, discrete phonon spectrum is obtained by using

the periodic boundary condition. In this case, the smallest $\omega$ attainable for the ZA modes is given by

$$\omega_{min} = \frac{\sqrt{\langle d_3 \rangle}}{L^2}. \tag{11}$$

When $L^2$ is large enough, the Debye model can still be applied, and eq. (10a) reduces to a finite value:

$$\Lambda_{Z1} = \frac{a^2 s_0 k_B T}{8\pi m \langle d_3 \rangle} \ln \left[ \frac{1 - e^{-\frac{\Theta_{ZA}}{T}}}{1 - e^{-\frac{\Theta_{min}}{T}}} \right] + \frac{\hbar a^2}{2m\sqrt{\langle d_3 \rangle}} \left( 1 - \frac{s_0}{8\pi L^2} \right), \tag{12}$$

where $\Theta_{min} = \frac{\hbar \omega_{min}}{k_B}$. Minimiztion of eq. (9) at given temperature determines the equilibrium lattice constant $a$. Within the SCH model, the solution is obtained in a self-consistent way using eqs. (9, 10b-c, 12), while the spontaneous curvature is determined by eq. (7).

When the independent standard deviations $\sqrt{\Lambda_{L1}}$, $\sqrt{\Lambda_{T1}}$, $\sqrt{\Lambda_{Z1}}$, and the change of equilibrium lattice constant from its ground state value $da = (a - a_0)$ are all small quantities compared with the ground state lattice constant, a two-step expansion method can be developed for all the terms with $\langle ... \rangle$. Application of the method to eq. (9) gives

$$f_{SCH} = (f_{SCH})_0 + A(T)\tilde{\kappa}^2 + B(T)\tilde{\kappa}^4, \tag{13}$$

where $(f_{SCH})_0$ denotes the free energy per atom for flat graphene with lattice constant $a_0$, and $A(T) = A_0 + \gamma_z T \ln \left[ \frac{1-e^{-\frac{\Theta_{out}}{T}}}{1-e^{-\frac{\Theta_{min}}{T}}} \right] + \gamma_{z2} T^2 \text{Log}^2 \left[ \frac{1-e^{-\frac{\Theta_{out}}{T}}}{1-e^{-\frac{\Theta_{min}}{T}}} \right] + \gamma_2 T^2 + \gamma_3 T^3$. Here $B(T)$, $A_0$, $\gamma_z$, $\gamma_{z2}$, $\gamma_2$ and $\gamma_3$ are coefficients explicitly determined by the ground state force constants. The two-step expansion method and the expressions of coefficients can be found in Supplementary Information. Through eq. (13), the Landau theory for the "*wrinkling transition*" (the name is used to distinguish itself from the crumpling transition in tethered surfaces[23]) of graphene is established, with $A(T)$ and $B$ analytically expressed in terms of the interatomic potential for carbon. For 1m² sized graphene, we have $A_0 = 0.087 \text{ev}$, $\gamma_z = -2.377 \times 10^{-4} \text{ev/K}$, $\gamma_{z2} = -2.584 \times 10^{-8} \text{ev/K}^2$, $\gamma_2 = 1.301 \times 10^{-6} \text{ev/K}^2$, $\gamma_3 = -6.011 \times 10^{-11} \text{ev/K}^3$, and $B = 58.88 \text{ev}$. Here higher order terms of temperature in the expressions of $A(T)$ and

$B$ are not shown. At given temperature, $f_{SCH}$ should be minimized in term of $\tilde{\kappa}$. Thus the value of $\tilde{\kappa}$ at equilibrium state is determined by $\frac{\partial f_{SCH}}{\partial \tilde{\kappa}} = 0, \frac{\partial^2 f_{SCH}}{\partial \tilde{\kappa}^2} > 0$, which gives

$$\tilde{\kappa}^2 = \begin{cases} -\frac{A(T)}{2B(T)}, & for\ A(T) < 0 \\ 0, & for\ A(T) \geq 0 \end{cases} \quad (14)$$

The critical temperature is obtained from $A(T) = 0$, and the phase transition is of second-order. This wrinkling transition is characterized by the appearance of spontaneous curvature, which changes the equilibrium configuration of graphene from a flat state to a rippling state. Yet, according to statistical mechanics the real configuration of the material is fluctuated around its equilibrium configuration and thus the real curvature of the graphene is strongly affected by thermal fluctuation, which can be written as

$$\langle \tilde{\kappa}^2 \rangle = \tilde{\kappa}^2 + \tilde{\kappa}_t^2 \approx \tilde{\kappa}^2 + \frac{4\Lambda_{Z1}}{a^2}, \quad (15)$$

where $\tilde{\kappa}_t$ is termed "*thermal curvature*" which characterizes the curvature due to thermal fluctuation. Variation of $\tilde{\kappa}$ and $\tilde{\kappa}_t$ with temperature for 1m² sized graphene predicted by the Landau theory and the SCH calculation is shown in Figure 1a. A wrinkling transition with critical temperature $T_c = 8.5K$ is obtained from the Landau theory, characterized by appearance of the order parameter $\tilde{\kappa}$. One exotic property of this phase transition is that the order parameter is zero at ground state and **increases with temperature**, contradicting the common belief that the order parameter increases monotonically in moving away from $T_c$ to 0K as predicted by the Landau theory for general second order phase transitions[44]. On the other hand, it is shown that for both models, the value of $\tilde{\kappa}_t$ is at least an order of magnitude larger than $\tilde{\kappa}$, for which the wrinkling transition characterized by appearance of $\tilde{\kappa}$ has a tiny effect on $\sqrt{\langle \tilde{\kappa}^2 \rangle}$. As the experimentally observed curvature is $\sqrt{\langle \tilde{\kappa}^2 \rangle}$, the wrinkling transition of graphene is hidden under the large out-of-plane undulation due to thermal fluctuation. Yet, since $\tilde{\kappa}$ reflects the spontaneous tendency of the system towards stability while $\tilde{\kappa}_t$ derives directly from thermal fluctuation, the ripples described by $\tilde{\kappa}$ should be "harder" than those described by $\tilde{\kappa}_t$. In other words, according to the theory of soft modes [33, 34], we believe that the actual configuration of graphene is determined by $\tilde{\kappa}$ and then amplified by $\tilde{\kappa}_t$.

We also observe in Figure 1a a discrepancy of $\tilde{\kappa}$ (and $\tilde{\kappa}_t$) predicted by the Landau theory and by SCH calculation. Below 750K, $\tilde{\kappa}$ (and $\tilde{\kappa}_t$) predicted by the Landau theory is always smaller, due to omission of higher order terms of fluctuation in the truncated expansion of all the terms with $\langle ... \rangle$. Above 750K, $\tilde{\kappa}$ predicted by the Landau theory exceeds $\tilde{\kappa}$ predicted by SCH calculation rapidly. This is because when constructing the Landau theory, the variances $\Lambda_{j1}, (j = L, T, Z)$ are overestimated since they are not calculated in a self-consistent way. A detailed analysis of the applicability of the Landau theory is provided in the Supplementary Information.

On the other hand, the phase transition point at 8.5K obtained in the Landau theory is blurred in the SCH calculation, where $\tilde{\kappa}$ is found to be 0.009 at 0K and thus 1m² sized graphene is already rippling at ground state according to the SCH calculation. This blurring can be attributed to omission of higher order effects of fluctuation in the Landau theory. Similar phenomenon has been treated before when discussing the crumpling transition[45].

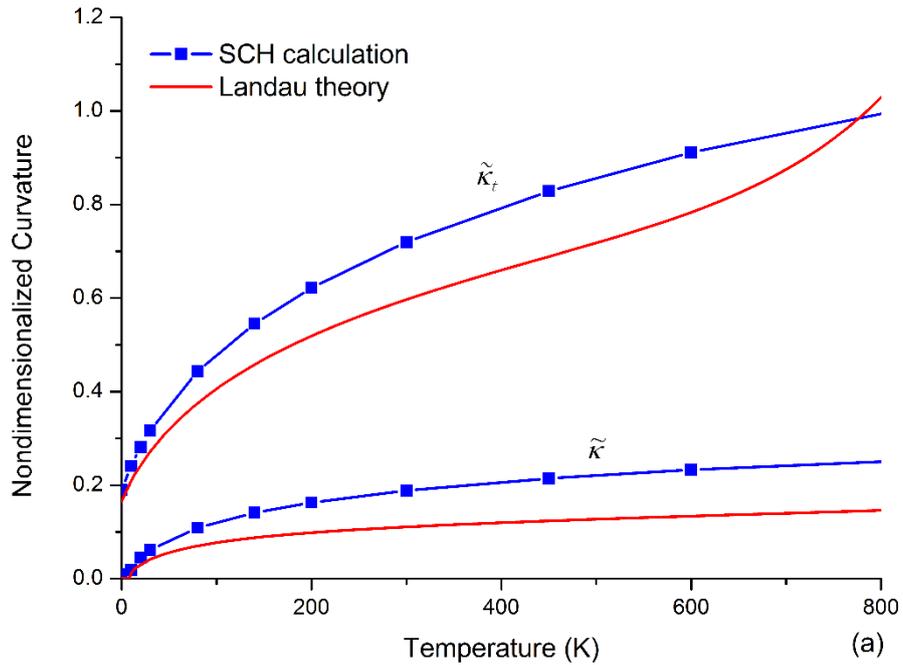

(a)

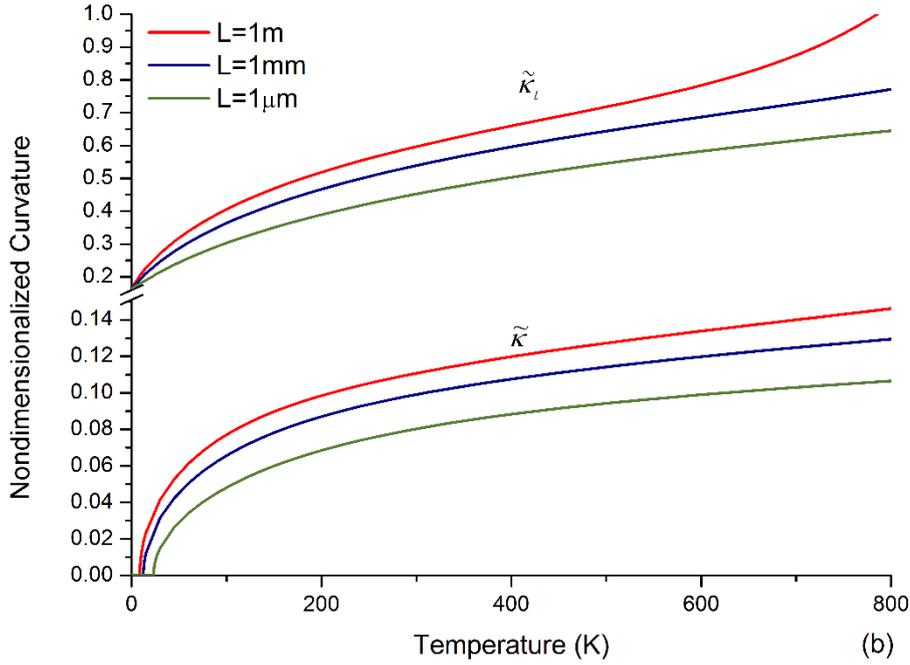

Figure 1. The variation of nondimensionalized curvature with temperature (a)using two different models and (b)for different sized graphene. For (a): the solid red curves plot the result of the Landau theory, while the blue lines with dots plot the result of SCH calculation. The variation of $\tilde{\kappa}_t$ is presented above the break region, and the variation of the spontaneous curvature $\tilde{\kappa}$ is presented below the break region. The size of graphene in this calculation is $1m^2$. For (b): results are obtained using the Landau theory, where the size of graphene is assumed to be $1m^2$ (red curve), $1mm^2$ (blue curve), and $1\mu m^2$ (green curve).

According to the Landau theory, the sign of $A(T)$ determines the occurrence of phase transition and the value of the order parameter. To further analyze the origin of the wrinkling transition, $A(T)$ in eq. (13) is recast into a new form: $A(T) = A_{QH} + A_L + A_T + A_Z + A_C$, where the explicit expression for $A_{QH}, A_L, A_T, A_Z$ and $A_C$ can be found in the Supplementary Information. Here $A_{QH}$ denotes the coefficient of $\tilde{\kappa}^2$ when the effect of thermal fluctuation is neglected in free energy calculation (i.e., $A_{QH}$ is obtained by using quasi-harmonic approximation); $A_L, A_T, A_Z$ and $A_C$ can be regarded as modification of $A_{QH}$ due to thermal vibration

of LA modes, TA modes, ZA modes, and coupling between these modes, respectively. For graphene, we have:

$A_{QH} = (0.31 + 1.301 \times 10^{-6} T^2 + 1.024 \times 10^{-9} T^3)$ev, $A_L = (-0.196 - 8.264 \times 10^{-10} T^3)$ev, $A_T = (0.017 - 2.195 \times 10^{-10} T^3)$ev, $A_Z = \left[ -0.044 - 2.291 \times 10^{-4} T \mathrm{Log}[\frac{1-e^{-\frac{\Theta_{out}}{T}}}{1-e^{-\frac{\Theta_{min}}{T}}}] - 2.584 \times 10^{-8} T^2 \mathrm{Log}^2[\frac{1-e^{-\frac{\Theta_{out}}{T}}}{1-e^{-\frac{\Theta_{min}}{T}}}] \right]$ev, and $A_C = (-0.0007 - 1.299 \times 10^{-6} T \mathrm{Log}[\frac{1-e^{-\frac{\Theta_{out}}{T}}}{1-e^{-\frac{\Theta_{min}}{T}}}])$ev.

The expressions of these coefficients are all composed of a constant term and some temperature-dependent terms. For $A_{QH}$, the constant term reflects the change of equilibrium potential energy, and the temperature-dependent terms derive from the energy of phonon vibration, where the $T^2$ term denotes the contribution from ZA modes and the $T^3$ term denotes the contribution from LA and TA modes. The positive of $A_{QH}$ at any temperature indicates that the appearance of $\tilde{\kappa}$ is energetically unfavorable within the quasi-harmonic approximation. Within the self-consistent phonon theory, however, the static atomistic potential is replaced by its ensemble average, or called the "smeared potential"[41], leading to a modification of $A_{QH}$ due to bond length fluctuation where $A_L, A_T, A_Z$ and $A_C$ are derived. The constant terms of $A_L, A_T, A_Z$ and $A_C$ represent the effect of zero-point fluctuation, which tends to stabilize $\tilde{\kappa}$ mainly through the vibration of LA modes. On the other hand, the temperature-dependent terms of $A_L, A_T, A_Z$ and $A_C$ all facilitate the appearance of $\tilde{\kappa}$ in graphene. Interestingly, due to the special $\omega \propto q^2$ form of dispersion relation for the ZA modes at long wavelength limit, $A_Z$ and $A_C$ can be regarded as linear function of temperature in the range $T \gg \Theta_{min}$. Besides, due to the strong structural anisotropy of graphene, the amplitude of out-of-plane vibration is significantly larger than the amplitude of in-plane vibration, for which the temperature-dependent term in $A_Z$ dominates in a wide temperature range up to around 1000K. At higher temperature, the negative $T^3$ terms in $A_L$ and $A_T$ offset the positive $T^3$ term in $A_{QH}$. Therefore, at low temperature a wrinkling transition takes place due to the negative temperature-dependent term in $A_Z$, while at finite temperature the equilibrium value of $\tilde{\kappa}$ is determined via the competition between negative temperature-dependent terms in $A_Z$ and the positive $T^2$ term in $A_{QH}$. This explains why the order parameter is stabilized rather than destroyed by an increase of temperature. Since NTE, large out-of-plane vibrational amplitude and the special $\omega \propto q^2$ form of dispersion relation for the ZA modes at long wavelength limit are three general properties for any 2D materials[24], the wrinkling transition and the associated heating-stabilized spontaneous curvature discussed here are of universal interest at least for tethered membranes.

Since $\Theta_{min}$ is determined by the size of graphene sample, the size effect on $\tilde{\kappa}_t$ and $\tilde{\kappa}$ is studied within eq. (13) and shown in Figure 1b. It is found that the critical temperature of the wrinkling transition is slightly increased when the graphene sample size drops from 1m² to 1mm² and then to 1μm². As for $\tilde{\kappa}_t$, the smaller the graphene sample size, the slower the variation of $\tilde{\kappa}_t$ with temperature.

From Eq. (14), we can express $\tilde{\kappa}^2$ as a function of $da$, for which eq. (13) can be reformulated as a function of $da$. Solving $\frac{\partial f_{SCH}}{\partial (da)} = 0$, we have

$$da = \begin{cases} \dfrac{C_{1l}}{C_{2l}}, & T \geq T_c \\ \dfrac{C_{1h}}{C_{2h}}, & T < T_c \end{cases} \quad (16)$$

where $C_{1l}, C_{2l}, C_{1l}$ and $C_{2l}$ are explicitly expressed in terms of the force constants in the Supplementary Information.

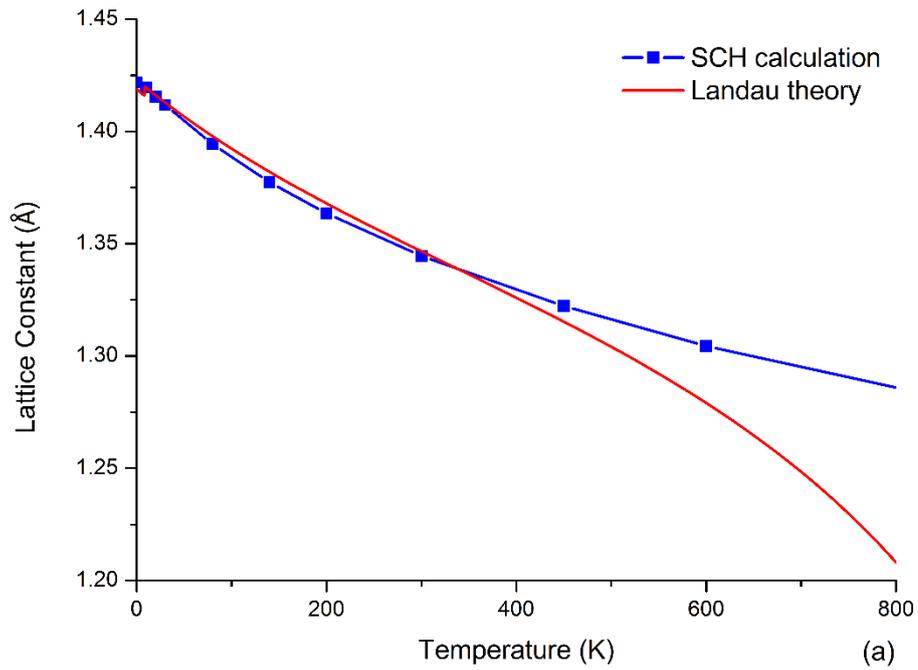

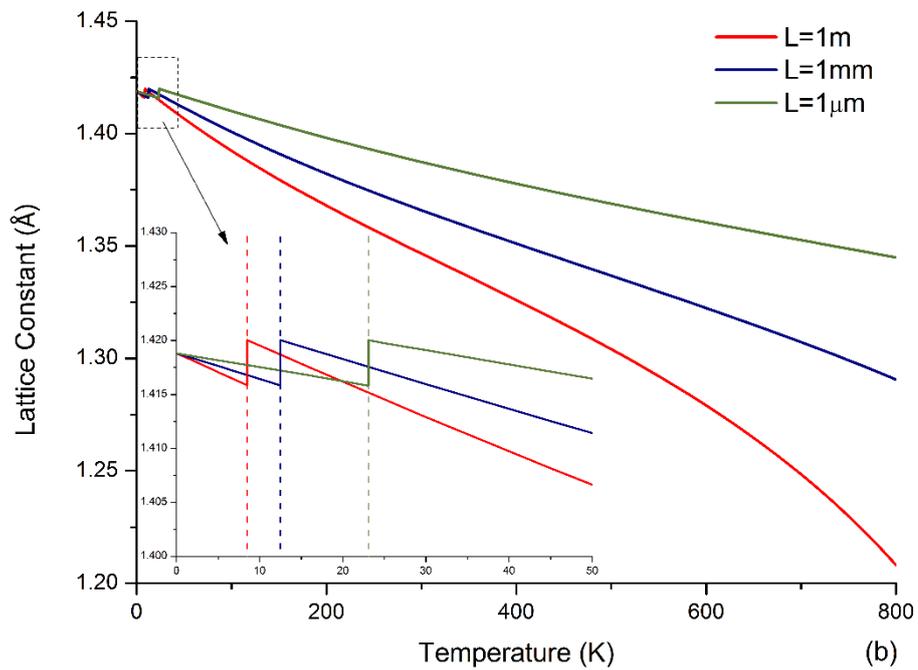

Figure 2. Variation of equilibrium lattice constant with temperature (a)using Landau theory (red curve) and SCH calculation (blue curve with dots), where the size of graphene in this calculation is 1m$^2$; (b)using Landau theory for discussion of the size effect, where the size of graphene is assumed to be 1m$^2$ (red curve), 1mm$^2$ (blue

curve), and 1μm² (green curve). In the inset, all dashed lines with different colors mark the critical temperature at which a sudden increase of lattice constant takes place.

The equilibrium lattice constant $a = a_0 + da$ as a function of temperature is plotted in Figure 2a using eq. (16) (the Landau theory) and the SCH calculation. It is observed that a sudden increase of equilibrium lattice constant at the critical temperature occurs according to the Landau theory, which is blurred out in SCH calculation due to higher order effect of fluctuation. For similar reasons shown in the discussion of $\tilde{\kappa}_t$ in Figure 1a, in Figure 2a the magnitude of NTE predicted by the Landau theory is slightly decreased compared with corresponding result from SCH calculation below around 340K, while above 340K, the prediction of the Landau theory exceeds that of the SCH calculation rapidly. We also observe in Figure 2b that NTE is weakened as the sample size drops from 1m² to 1mm² and then to 1μm².

If we describe graphene as a thin plate, the Helmholtz free energy density can be obtained from eq. (13) as $F_{SCH} = \frac{2f_{SCH}}{s_0}$. According to the elastic theory of thin plates[46], the equibiaxial bending moment $M = M_{11} = M_{22}$ and the equibiaxial bending curvature $\kappa = \kappa_{11} = \kappa_{22}$ forms a conjugate pair when concerning the strain energy density of thin plate. For a local region in graphene with spontaneous curvature $\tilde{\kappa}$, we define the spontaneous bending rigidity as $\widetilde{D} = \frac{\partial^2 F_{SCH}}{\partial \kappa^2}$ where $\kappa = \tilde{\kappa}/a$. From eq. (13) we have $\widetilde{D} = \frac{8\sqrt{3}}{9}A$ for $T \geq T_c$ and $\widetilde{D} = -\frac{16\sqrt{3}}{9}A$ for $T < T_c$. we also define the thermal bending rigidity as $D = \frac{\partial^2 F_{SCH}}{\left(\partial\sqrt{\langle\kappa^2\rangle}\right)^2}$, where $\langle\kappa^2\rangle = \langle\tilde{\kappa}^2\rangle/a^2$. As shown in Figure 3, the wrinkling transition is accompanied by vanishing of $\widetilde{D}$, which, however, is hidden under the strong effect of thermal fluctuation. Because we can see that the effect of vanishing of $\widetilde{D}$ is almost unobservable in the curve of $D$, which is at least two orders of magnitude larger than $\widetilde{D}$. This result explains the large discrepancy between theoretical prediction of the bending rigidity of suspended graphene [15, 32, 47] (which is close to $\widetilde{D}$) and corresondng experimental value obtained recently[48] (which is close to $D$). One should keep in mind that our theory, established upon the Debye model, is constructed for large sized graphene. When applying it to small sized sample, the zero-point fluctuation is overestimated using the Debye model, and so the value of $D$ predicted here should be regarded as an upper bound for finite sized graphene.

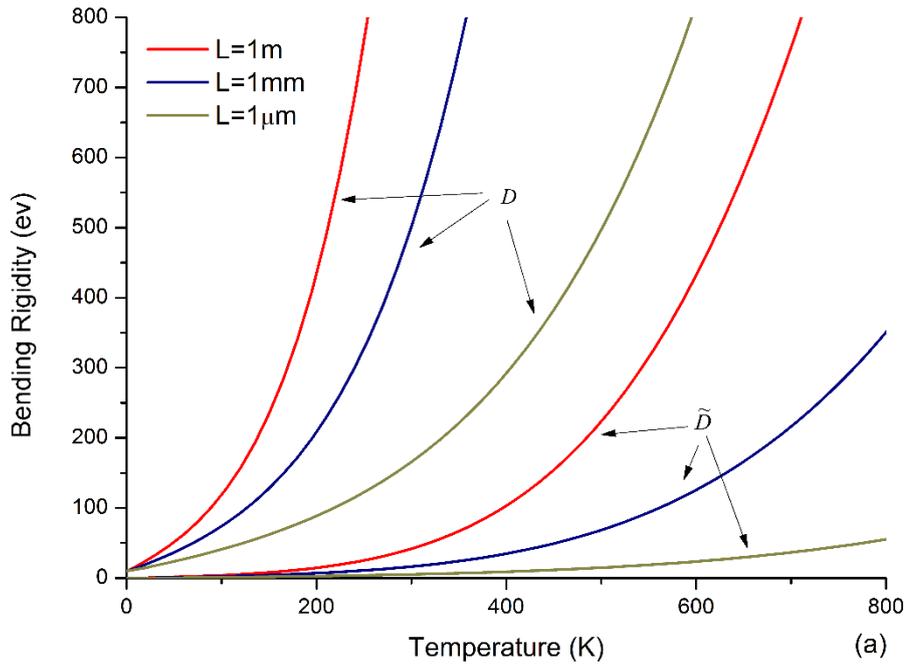

(a)

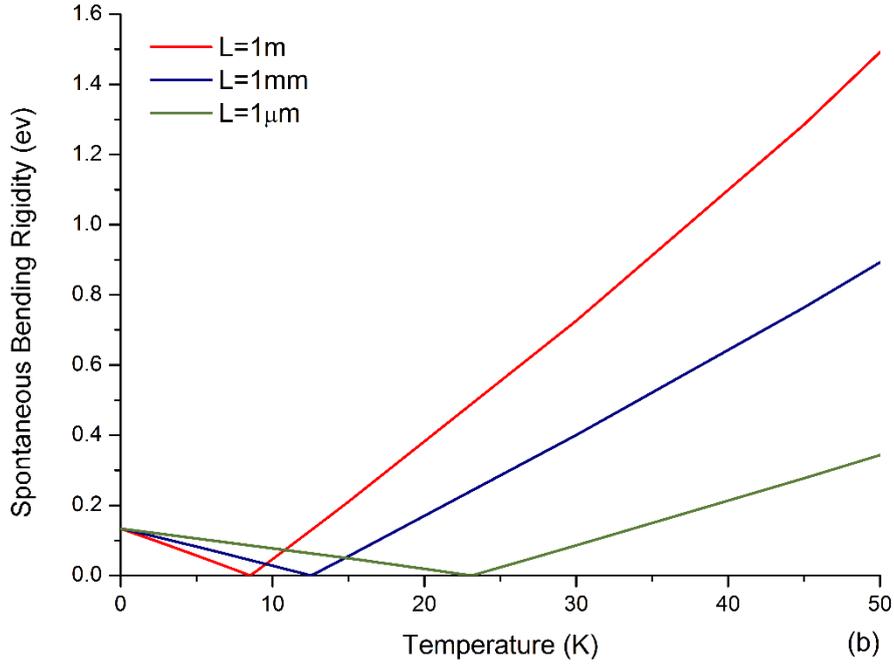

Figure 3. Variation of (a) the spontaneous bending rigidity $\widetilde{D}$ (below the break region) and the thermal bending rigidity $D$ (above the break region) with temperature (below 800K) for graphene sample with different size; (b) the spontaneous bending rigidity $\widetilde{D}$ with temperature (near the phase transition point) for graphene sample with different size

Through analyzing the wrinkling transition of suspended graphene based on the established Landau theory, it is verified that interplay between the equilibrium morphology and soft ZA phonon modes in 2D materials is vital for their stability. According to the theory of soft modes, the slow varying modes determine the long-term behavior of the system[49]. The results obtained not only establish a foundation for further research on the morphology evolution of 2D materials exposed to various external fields, but also promote studies on the coupling between the ZA phonon mode and other elementary excitations, which is significant for understanding the properties of actual 2D materials and for exploring the possibility of ripple-engineering for 2D materials. The result that spontaneous curvature may exist for any tethered membranes implies soft mode dynamics may be the key to understanding the widely observed rippling phenomena in membrane-like materials.

1. Meyer JC, Geim AK, Katsnelson MI, Novoselov KS, Booth TJ, Roth S. The structure of suspended graphene sheets. *Nature* 2007, **446**(7131)**:** 60-63.

2. Chen CC, Bao WZ, Theiss J, Dames C, Lau CN, Cronin SB. Raman Spectroscopy of Ripple Formation in Suspended Graphene. *Nano letters* 2009, **9**(12)**:** 4172-4176.

3. Tapaszto L, Dumitrica T, Kim SJ, Nemes-Incze P, Hwang C, Biro LP. Breakdown of continuum mechanics for nanometre-wavelength rippling of graphene. *Nat Phys* 2012, **8**(10)**:** 739-742.

4. Landau L. ZhETF 7 819 (1937); Phys. *Z Sowjet* 1937, **11:** 556.

5. Mermin N. Crystalline Order in Two Dimensions. *Physical Review* 1968, **176**(1)**:** 250-254.

6. Nelson D, Peliti L. Fluctuations in membranes with crystalline and hexatic order. *Journal de physique* 1987, **48**(7)**:** 1085-1092.

7. Brivio J, Alexander DT, Kis A. Ripples and layers in ultrathin MoS2 membranes. *Nano letters* 2011, **11**(12)**:** 5148-5153.

8. Jose D, Datta A. Structures and electronic properties of silicene clusters: a promising material for FET and hydrogen storage. *Phys Chem Chem Phys* 2011, **13**(16)**:** 7304-7311.

9. Elias D, Nair R, Mohiuddin T, Morozov S, Blake P, Halsall M*, et al.* Control of graphene's properties by reversible hydrogenation: evidence for graphane. *Science* 2009, **323**(5914)**:** 610-613.

10. Singh SK, Neek-Amal M, Costamagna S, Peeters FM. Thermomechanical properties of a single hexagonal boron nitride sheet. *Physical Review B* 2013, **87**(18).

11. Tombros N, Jozsa C, Popinciuc M, Jonkman HT, van Wees BJ. Electronic spin transport and spin precession in single graphene layers at room temperature. *Nature* 2007, **448**(7153)**:** 571-U574.

12. Okada Y, Zhou WW, Walkup D, Dhital C, Wilson SD, Madhavan V. Ripple-modulated electronic structure of a 3D topological insulator. *Nat Commun* 2012, **3**.

13. Simon F, Muranyi F, Dora B. Theory and model analysis of spin relaxation time in graphene - Could it be used for spintronics? *Phys Status Solidi B* 2011, **248**(11)**:** 2631-2634.

14. Lee C, Wei X, Kysar JW, Hone J. Measurement of the elastic properties and intrinsic strength of monolayer graphene. *Science* 2008, **321**(5887)**:** 385-388.

15. Zhang D-B, Akatyeva E, Dumitrică T. Bending ultrathin graphene at the margins of continuum mechanics. *Physical review letters* 2011, **106**(25): 255503.

16. Bunch JS, van der Zande AM, Verbridge SS, Frank IW, Tanenbaum DM, Parpia JM, *et al.* Electromechanical resonators from graphene sheets. *Science* 2007, **315**(5811): 490-493.

17. Liu L, Ryu S, Tomasik MR, Stolyarova E, Jung N, Hybertsen MS, *et al.* Graphene oxidation: thickness-dependent etching and strong chemical doping. *Nano letters* 2008, **8**(7): 1965-1970.

18. Boukhvalov DW, Katsnelson MI. Enhancement of Chemical Activity in Corrugated Graphene. *J Phys Chem C* 2009, **113**(32): 14176-14178.

19. Weeks JD. The roughening transition. *Ordering in strongly fluctuating condensed matter systems*. Springer, 1980, pp 293-317.

20. Zasadzinski J, Schneir J, Gurley J, Elings V, Hansma PK. Scanning tunneling microscopy of freeze-fracture replicas of biomembranes. *Science* 1988, **239**(4843): 1013-1015.

21. Antonietti M, Wenzel A, Thünemann A. The "egg-carton" phase: A new morphology of complexes of polyelectrolytes with natural lipid mixtures. *Langmuir* 1996, **12**(8): 2111-2114.

22. Fendler JH, Tundo P. Polymerized surfactant aggregates: characterization and utilization. *Accounts Chem Res* 1984, **17**(1): 3-8.

23. Nelson D. *Statistical mechanics of membranes and surfaces*. World Scientific, 2004.

24. Hu YF, Chen JP, Wang B. On the Intrinsic Ripples and Negative Thermal Expansion of Graphene. *submitted* 2015.

25. Paczuski M, Kardar M, Nelson DR. Landau theory of the crumpling transition. *Physical review letters* 1988, **60**(25): 2638.

26. Kosterlitz J, Thouless D. Long range order and metastability in two dimensional solids and superfluids.(Application of dislocation theory). *Journal of Physics C: Solid State Physics* 1972, **5**(11): L124.

27. Halperin B, Nelson D. Theory of Two-Dimensional Melting. *Physical Review Letters* 1978, **41**(2): 121-124.

28. Nelson DR, Halperin B. Dislocation-mediated melting in two dimensions. *Physical Review B* 1979, **19**(5): 2457.

29. Young A. Melting and the vector Coulomb gas in two dimensions. *Physical Review B* 1979, **19**(4): 1855-1866.

30. Los JH, Ghiringhelli LM, Meijer EJ, Fasolino A. Improved long-range reactive bond-order potential for carbon. I. Construction. *Physical Review B* 2005, **72**(21).

31. Mariani E, von Oppen F. Flexural phonons in free-standing graphene. *Physical review letters* 2008, **100**(7): 076801.

32. Fasolino A, Los JH, Katsnelson MI. Intrinsic ripples in graphene. *Nature materials* 2007, **6**(11): 858-861.

33. Lines ME, Glass AM. *Principles and applications of ferroelectrics and related materials*. Clarendon press Oxford, 2001.

34. Blinc R, Žekš B. *Soft modes in ferroelectrics and antiferroelectrics*. North-Holland Publishing Company Amsterdam, 1974.

35. Kadanoff LP. Scaling laws for Ising models near Tc. *Physics* 1966, **2**(6): 263-272.

36. San-Jose P, González J, Guinea F. Electron-induced rippling in graphene. *Physical review letters* 2011, **106**(4): 045502.

37. Maradudin AA, Montroll EW, Weiss GH, Ipatova I. *Theory of lattice dynamics in the harmonic approximation*, vol. 3. Academic press New York, 1963.

38. Stewart GW, Sun J-g, Jovanovich HB. *Matrix perturbation theory*, vol. 175. Academic press New York, 1990.

39. Kosevich AM. *Front Matter*. Wiley Online Library, 2005.

40. Koehler T. Theory of the Self-Consistent Harmonic Approximation with Application to Solid Neon. *Physical Review Letters* 1966, **17**(2): 89-91.

41. Choquard P. *The anharmonic crystal*. WA Benjamin New York, 1967.

42. Werthamer N. Self-Consistent Phonon Formulation of Anharmonic Lattice Dynamics. *Physical Review B* 1970, **1**(2): 572-581.

43. Landau LD, Lifshitz E. Statistical Physics. Part 1: Course of Theoretical Physics. Pergamon; 1980.

44. Izi͡umov IUrA, Syromyatnikov VN. *Phase transitions and crystal symmetry*, vol. 38. Springer Science & Business Media, 1990.

45. De Gennes P-G. *Scaling concepts in polymer physics*. Cornell university press, 1979.

46. Reddy JN. *Theory and analysis of elastic plates and shells*. CRC press, 2006.

47. Lindahl N, Midtvedt D, Svensson J, Nerushev OA, Lindvall N, Isacsson A*, et al.* Determination of the Bending Rigidity of Graphene via Electrostatic Actuation of Buckled Membranes. *Nano letters* 2012, **12**(7)**:** 3526-3531.

48. Nicholl R, Conely H, Lavrik N, Vlassiouk I, Bolotin K. Probing Mechanics of Rippled Two-Dimensional Materials. *Bulletin of the American Physical Society* 2015, **60**.

49. Haken H. *Synergetics*. Springer, 1977.

# Supplementary Information

## 1. Perturbational phonon dispersion analysis for rippled graphene

According to the theory of lattice dynamics in the harmonic approximation[s1], the phonon frequencies for any crystalline solid at low temperature are determined by

$$\omega_{\mathbf{q}n}^2 = \sum_{\alpha,\beta,k,k'} \varepsilon_{\alpha k}(\mathbf{q},n)\varepsilon_{\beta k'}(\mathbf{q},n) D_{\alpha\beta}\begin{pmatrix}\mathbf{q}\\k,k'\end{pmatrix}, \qquad (S1)$$

where $\varepsilon_{\alpha k}(\mathbf{q},n)$ is polarization of the $k$th atom in the unit cell in the $\alpha$-axis direction of the $n$th branch of phonon with wave vector $\mathbf{q}$, and

$$D_{\alpha\beta}\begin{pmatrix}\mathbf{q}\\k,k'\end{pmatrix} = \frac{1}{m}\sum_{\mathbf{l}'} K_{\alpha\beta}\begin{pmatrix}\mathbf{0},\mathbf{l}'\\k,k'\end{pmatrix} e^{-i\mathbf{q}\mathbf{l}'_{k'}} \qquad (S2)$$

denote the components of the dynamical matrix. $\mathbf{l}'_{k'}$ is the lattice vector pointing to the unit cell in which atom $k'$ is lying. In eq. (2)

$$K_{\alpha\beta}\begin{pmatrix}\mathbf{0},\mathbf{l}'\\kk'\end{pmatrix} = \left(\frac{\partial^2 \mathrm{U}}{\partial u_\alpha\begin{pmatrix}\mathbf{0}\\k\end{pmatrix}\partial u_\beta\begin{pmatrix}\mathbf{l}'\\k'\end{pmatrix}}\right)_{r=a}, \qquad (S3)$$

where $\mathrm{U} = \sum_i \Phi_i$ denotes the total potential energy of the system, and $\Phi_i$ is the potential energy possessed by the $i$th atom. $u_\alpha\begin{pmatrix}\mathbf{0}\\k\end{pmatrix}$ denotes the $\alpha$th component of the displacement vector of the $k$th atom in the $\mathbf{0}$th unit cell. For graphene described by LCBOP II [s2], $\Phi_i$ takes the following form

$$\Phi_i = \frac{1}{2}\left[\sum_{j,|\mathbf{r}_{ij}|=a} \varphi_1(\mathbf{r}_{ij}, \theta_{ijk}) + \sum_{j,|\mathbf{r}_{ij}|=\sqrt{3}a} \varphi_2(\mathbf{r}_{ij}) + \sum_{j,|\mathbf{r}_{ij}|=2a} \varphi_3(\mathbf{r}_{ij})\right], \quad (S4)$$

where $\theta_{ijk}$ denotes the bond angle between neighboring bond $\mathbf{r}_{ij}$ and $\mathbf{r}_{ik}$ ($k \neq i,j$). $\varphi_1(\mathbf{r}_{ij}, \theta_{ijk})$, $\varphi_2(\mathbf{r}_{ij})$ and $\varphi_3(\mathbf{r}_{ij})$ denote respectively the nearest neighbor interaction, the second nearest neighbor interaction and the third nearest neighbor interaction. Explicit expressions of the phonon frequencies $\omega_{\mathbf{q}n}$ are usually difficult to obtain due to the complex expression of $D_{\alpha\beta}\begin{pmatrix}\mathbf{q}\\k,k'\end{pmatrix}$. Yet, at long wavelength limit (i.e., for small $q = |\mathbf{q}|$), the analytical phonon dispersion relation can be obtained by using the matrix perturbation theory[s3]. The methodology is briefly reviewed as follow: $\mathbf{Q} = \mathbf{Q}_0 + \varepsilon\mathbf{Q}_1 + \varepsilon^2\mathbf{Q}_2$ is a target matrix where $\varepsilon$ is considered small perturbation, and $\mathbf{Q}_0$ is the unperturbed matrix which satisfies

$$\begin{aligned}\mathbf{Q}_0\mathbf{u}_0 &= \mathbf{u}_0\mathbf{\Lambda}_0,\\ \mathbf{u}_0^T\mathbf{u}_0 &= \mathbf{I}.\end{aligned} \quad (S5)$$

Here $\mathbf{u}_0$ is a matrix whose columes are the unit normal eigenvectors of $\mathbf{Q}_0$, $\mathbf{\Lambda}_0$ is the diagonalized matrix of $\mathbf{Q}_0$, and $\mathbf{I}$ denotes the identity matrix of the same order with $\mathbf{\Lambda}_0$. Consider the case where $\mathbf{Q}_0$ has an eigenvalue with multiplicity $m$, denoted by $\lambda_0 = \lambda_{01} = \lambda_{02} = \cdots \lambda_{0m}$, and we are particularly interested in the perturbation of $\lambda_{01}, \lambda_{02}, \ldots, \lambda_{0m}$ by presence of $\mathbf{Q}_1$ and $\mathbf{Q}_2$. The perturbed eigenvalues take the following form:

$$\mathbf{\Lambda}_m = \mathbf{\Lambda}_{0m} + \varepsilon\mathbf{\Lambda}_{1m} + \varepsilon^2\mathbf{\Lambda}_{2m}, \quad (S6)$$

where $\mathbf{\Lambda}_{0m} = \lambda_0\mathbf{I}_m$ and $\mathbf{I}_m$ denotes the identity matrix of order $m$. After manipulation we have

$$\mathbf{u}_{0m}^T\mathbf{Q}_1\mathbf{u}_{0m}\boldsymbol{\alpha} = \boldsymbol{\alpha}\mathbf{\Lambda}_{1m}. \quad (S7)$$

If $\mathbf{u}_{0m}^T\mathbf{Q}_1\mathbf{u}_{0m}$ equals to a zero matrix, we have

$$(-\mathbf{u}_{0m}^T\mathbf{Q}_1\mathbf{u}_{0a}(\mathbf{\Lambda}_{0a} - \lambda_0\mathbf{I}_a)^{-1}\mathbf{u}_{0a}^T\mathbf{Q}_1\mathbf{u}_{0m} + \mathbf{u}_{0m}^T\mathbf{Q}_2\mathbf{u}_{0m})\boldsymbol{\alpha} = \boldsymbol{\alpha}\mathbf{\Lambda}_{2m}. \quad (S8)$$

Here $\mathbf{u}_{0m}$ is a matrix whose columes are the unit normal eigenvectors corresponding to $\lambda_{01}, \lambda_{02}, \ldots, \lambda_{0m}$; $\mathbf{\Lambda}_{0a}$ is a diagonal matrix composed by the eigenvalues of $\mathbf{Q}_0$ excluding $\lambda_{01}, \lambda_{02}, \ldots, \lambda_{0m}$; $\mathbf{u}_{0a}$ is a matrix whose columes are the unit normal eigenvectors corresponding to $\mathbf{\Lambda}_{0a}$. Eqs. (S7, S8) give the eigen-equations for $\mathbf{\Lambda}_{1m}$ and $\mathbf{\Lambda}_{2m}$, which determine the perturbation of $\lambda_{01}, \lambda_{02}, \ldots, \lambda_{0m}$. It is obvious that the case of isolated eigenvalues can also be treated if we let $m = 1$. Eqs. (S5-S8) provide all the results we need for matrix perturbation analysis of the dispersion relation of graphene to the lowest order. At long wavelength limit, for flat graphene we have from eq. (2)

$$\mathbf{D}_0\begin{pmatrix}\mathbf{q}\\k,k'\end{pmatrix} = \mathbf{D}_0\begin{pmatrix}\mathbf{0}\\k,k'\end{pmatrix} + \mathbf{D}_{01}\begin{pmatrix}\mathbf{0}\\k,k'\end{pmatrix}q + \mathbf{D}_{02}\begin{pmatrix}\mathbf{0}\\k,k'\end{pmatrix}q^2, \tag{S9}$$

where $\mathbf{D}_{01}$ and $\mathbf{D}_{02}$ denote respectively the first and second order expansion coefficient matrix of $\mathbf{D}_0$ in terms of $q$. Using the method developed through eqs. (S5-S8), the dispersion relation for the acoustic phonon branches of flat graphene at long wavelength limit can be obtained as

$$\omega_{ZA}^2 = b_3 q^2 + d_3 q^4, \tag{S10a}$$

$$\omega_{LA}^2 = b_1 q^2, \tag{S10b}$$

$$\omega_{TA}^2 = b_2 q^2, \tag{S10c}$$

where

$$b_1 = \frac{1}{m}\left(-\frac{a_2^2}{2a_1} + a_3 + a_4\right), \tag{S11a}$$

$$b_2 = \frac{1}{m}\left(-\frac{a_2^2}{2a_1} + a_5 + a_6\right), \tag{S11b}$$

$$b_3 = \frac{3}{4m}a(c_{11} + 2\sqrt{3}c_{21} + 2c_{31}), \tag{S11c}$$

$$d_3 = \frac{3}{64m}a^2[24c_a - a(c_{11} + 6\sqrt{3}c_{21} + 8c_{31})], \tag{S11d}$$

and

$$a_1 = \frac{3(2a^2(c_{12} + c_{32}) + 3(4c_a + 6c_{a2} - 3c_{ab}) + a(2c_{11} + c_{31} + 12c_{ar}))}{4a^2}, \tag{S12a}$$

$$a_2 = \frac{3(-2a(c_{11} - c_{31}) + 2a^2(c_{12} - 2c_{32}) - 3(4c_a + 6c_{a2} - 3c_{ab}))}{8a}, \tag{S12b}$$

$$a_3 = \frac{3}{8}a(\sqrt{3}c_{21} + 9ac_{22} - 9c_{ar}), \tag{S12c}$$

$$a_4 = \frac{3}{32}\left(6a^2(c_{12} + 4c_{32}) + 3(4c_a + 6c_{a2} - 3c_{ab}) + 2a(c_{11} + 2(c_{31} + 6c_{ar}))\right), \tag{S12d}$$

$$a_5 = \frac{9}{16}(2\sqrt{3}ac_{21} + 2a^2c_{22} - 4c_a - 6c_{a2} + 3c_{ab} - 6ac_{ar}), \tag{S12e}$$

$$a_6 = \frac{3}{32}(2a^2(c_{12} + 4c_{32}) + 9(4c_a + 6c_{a2} - 3c_{ab}) + 6a(c_{11} + 2c_{31} + 4c_{ar})). \tag{S12f}$$

The force constants that appear in eqs. (12a-f) are defined by

$$c_{11} = \left(\frac{\partial \varphi_1}{\partial r_{ij}}\right)_0, \quad c_{21} = \left(\frac{\partial \varphi_2}{\partial r_{ij}}\right)_0, \quad c_{31} = \left(\frac{\partial \varphi_3}{\partial r_{ij}}\right)_0,$$

$$c_{12} = \left(\frac{\partial^2 \varphi_1}{\partial r_{ij}^2}\right)_0, \quad c_{22} = \left(\frac{\partial^2 \varphi_2}{\partial r_{ij}^2}\right)_0, \quad c_{32} = \left(\frac{\partial^2 \varphi_3}{\partial r_{ij}^2}\right)_0,$$

$$c_a = \left(\frac{\partial \varphi_1}{\partial \cos \theta_{ijk}}\right)_0, \quad c_{a2} = \left(\frac{\partial^2 \varphi_1}{\partial \cos \theta_{ijk} \, \partial \cos \theta_{ijk}}\right)_0,$$

$$c_{ab} = \left(\frac{\partial^2 \varphi_1}{\partial \cos \theta_{ijk} \, \partial \cos \theta_{ijl}}\right)_0, \quad c_{ar} = \left(\frac{\partial^2 \varphi_1}{\partial \cos \theta_{ijk} \, \partial r_{ij}}\right)_0.$$

The geometry of rippling graphene can be assumed as being composed of many local regions with homogeneous curvature. For a local region with nondimensionalized curvature $\bar{\kappa}$, the dynamical matrix can be expanded as

$$\mathbf{D}\begin{pmatrix}\mathbf{q}\\k,k'\end{pmatrix} = \mathbf{D}_0\begin{pmatrix}\mathbf{q}\\k,k'\end{pmatrix} + \mathbf{D}_1\begin{pmatrix}\mathbf{q}\\k,k'\end{pmatrix}\bar{\kappa} + \mathbf{D}_2\begin{pmatrix}\mathbf{q}\\k,k'\end{pmatrix}\bar{\kappa}^2. \quad (S13)$$

For different local regions, $\bar{\kappa}$ can take different values and different signs. Yet, due to the absence of macroscopic bending moment, the average of $\bar{\kappa}$ over the whole material should be zero and the neighboring regions of one local region with positive $\bar{\kappa}$ should be negative to ensure equilibrium of local bending moment. At equilibrium $|\bar{\kappa}|$ of an arbitrary local region would fluctuate around $\tilde{\kappa}$. Thus for a local region with positive $\bar{\kappa}$ (or called a local crest), eq. (S13) can be approximated by

$$\mathbf{D}^{crest}\begin{pmatrix}\mathbf{q}\\k,k'\end{pmatrix} = \mathbf{D}_0\begin{pmatrix}\mathbf{q}\\k,k'\end{pmatrix} + \mathbf{D}_1\begin{pmatrix}\mathbf{q}\\k,k'\end{pmatrix}\tilde{\kappa} + \mathbf{D}_2\begin{pmatrix}\mathbf{q}\\k,k'\end{pmatrix}\tilde{\kappa}^2. \quad (S14)$$

For its neighboring regions ( called local troughs)

$$\mathbf{D}^{trough}\begin{pmatrix}\mathbf{q}\\k,k'\end{pmatrix} = \mathbf{D}_0\begin{pmatrix}\mathbf{q}\\k,k'\end{pmatrix} - \mathbf{D}_1\begin{pmatrix}\mathbf{q}\\k,k'\end{pmatrix}\tilde{\kappa} + \mathbf{D}_2\begin{pmatrix}\mathbf{q}\\k,k'\end{pmatrix}\tilde{\kappa}^2. \quad (S15)$$

At long wavelength limit, we are interested in the vibrational displacement of a local crest as a whole, denoted by $\mathbf{u}_{crest}$, and the vibrational displacement of one neighboring local trough as a whole, denoted by $\mathbf{u}_{trough}$. An alternative choice is to study $\frac{1}{2}(\mathbf{u}_{crest} + \mathbf{u}_{trough})$ and $\frac{1}{2}(\mathbf{u}_{crest} - \mathbf{u}_{trough})$, which at long wavelength limit and for big enough material size denote the uniform motion and relative motion of one local crest and one neighboring local trough. Since soft modes appear among ZA modes, we are perticularly interested in $\frac{1}{2}(\mathbf{u}_{crest} + \mathbf{u}_{trough})$, whose dynamic matrix can be approximated by

$$\mathbf{D}^{acoustic}\begin{pmatrix}\mathbf{q}\\k,k'\end{pmatrix} = \mathbf{D}_0\begin{pmatrix}\mathbf{q}\\k,k'\end{pmatrix} + \mathbf{D}_2\begin{pmatrix}\mathbf{q}\\k,k'\end{pmatrix}\tilde{\kappa}^2. \tag{7}$$

For rippled graphene, the dynamic matrix for the acoustic branches at long wavelength limit is affected by the spontaneous curvature $\tilde{\kappa}$, as described by eq. (2) in the main text. At long wavelength limit we have

$$\begin{aligned}\mathbf{D}^{acoustic}\begin{pmatrix}\mathbf{q}\\k,k'\end{pmatrix} &= \mathbf{D}_0\begin{pmatrix}\mathbf{0}\\k,k'\end{pmatrix} + \mathbf{D}_{01}\begin{pmatrix}\mathbf{0}\\k,k'\end{pmatrix}q + \mathbf{D}_{02}\begin{pmatrix}\mathbf{0}\\k,k'\end{pmatrix}q^2 \\ &+ \left[\mathbf{D}_2\begin{pmatrix}\mathbf{0}\\k,k'\end{pmatrix} + \mathbf{D}_{21}\begin{pmatrix}\mathbf{0}\\k,k'\end{pmatrix}q + \mathbf{D}_{22}\begin{pmatrix}\mathbf{0}\\k,k'\end{pmatrix}q^2\right]\tilde{\kappa}^2.\end{aligned} \tag{S16}$$

We consider $\tilde{\kappa}^2$ as a perturbation, where the unperturbed eigenvalues are approximated by eqs. (S10a-c). Through the method in eqs. (S5-S8), the dispersion relation for the acoustic phonon branches perturbed by spontaneous curvature can be obtained as

$$\omega_{ZA}^2 = b_3^* q^2 + d_3 q^4, \quad b_3^* = b_3 + h_3(\alpha)\tilde{\kappa}^2, \tag{S17a}$$

$$\omega_{LA}^2 = b_1^* q^2, \quad b_1^* = b_1 + h_1(\alpha)\tilde{\kappa}^2, \tag{S17b}$$

$$\omega_{TA}^2 = b_2^* q^2, \quad b_2^* = b_2 + h_2(\alpha)\tilde{\kappa}^2, \tag{S17c}$$

where

$$\begin{aligned}h_3(\alpha) = -\frac{3}{256m}\{&2[-12a^2(c_{12}+14c_{22}+12c_{32}) - 9(12c_a+26c_{a2}-13c_{ab}) \\ &+4a(3c_{11}+14\sqrt{3}c_{21}+18c_{31}+15c_{ar})] + [12a^2(c_{12}-2(c_{22}+6c_{32})) \\ &-54(7c_a-2c_{a2}+c_{ab}) + a(-13c_{11}+10\sqrt{3}c_{21}+82c_{31}+120c_{ar})]\cos[2\alpha]\},\end{aligned} \tag{S18a}$$

$$\begin{aligned}h_1(\alpha) = -\frac{3}{1024m}\{&4[6a^2(c_{12}+14c_{22}+12c_{32}) + 9(16c_a+26c_{a2}-13c_{ab}) \\ &-2a(3c_{11}+14\sqrt{3}c_{21}+18c_{31}+24c_{ar})] + 2[6a^2(c_{12}+2c_{22}-4c_{32}) \\ &-9(2c_a-30c_{a2}+15c_{ab}) - a(4c_{11}+8\sqrt{3}c_{21}+8c_{31}-63c_{ar})]\cos[2\alpha] \\ &+[6a^2(3c_{12}+2c_{22}-28c_{32}) - 3(134c_a-90c_{a2}+45c_{ab}) \\ &+a(-19c_{11}-6\sqrt{3}c_{21}+94c_{31}+204c_{ar})]\cos[4\alpha]\},\end{aligned} \tag{S18b}$$

$$\begin{aligned}h_2(\alpha) = \frac{3}{1024m}\{&-12[6a^2(c_{12}+14c_{22}+12c_{32}) + 3(40c_a+18c_{a2}-9c_{ab}) \\ &-2a(3c_{11}+14\sqrt{3}c_{21}+18c_{31}+28c_{ar})] - 2[6a^2(3c_{12}-10c_{22}-44c_{32}) \\ &-47c_a+54c_{a2}-27c_{ab} + a(-18c_{11}+20\sqrt{3}c_{21}+132c_{31}+129c_{ar})]\cos[2\alpha] \\ &+[6a^2(3c_{12}+2c_{22}-28c_{32}) - 3(134c_a-90c_{a2}+45c_{ab}) \\ &+a(-19c_{11}-6\sqrt{3}c_{21}+94c_{31}+204c_{ar})]\cos[4\alpha]\}.\end{aligned} \tag{S18c}$$

From eqs. (S17a-c) we learn that the coefficients of the $q^2$ term in the dispersion relations for the acoustic phonon branches are renormalized by presence of the spontaneous curvature $\tilde{\kappa}$. Since $h_1, h_2$ and $h_3$ are $\alpha$ dependent while $b_1, b_2$ and $b_3$ are independent of $\alpha$, we can conclude that presence of $\tilde{\kappa}$ reduces the symmetry of the dispersion relations at long wavelength limit. Figure S1 shows the plot of $h_1, h_2$ and $h_3$ as functions of $\alpha$ in polar coordinate.

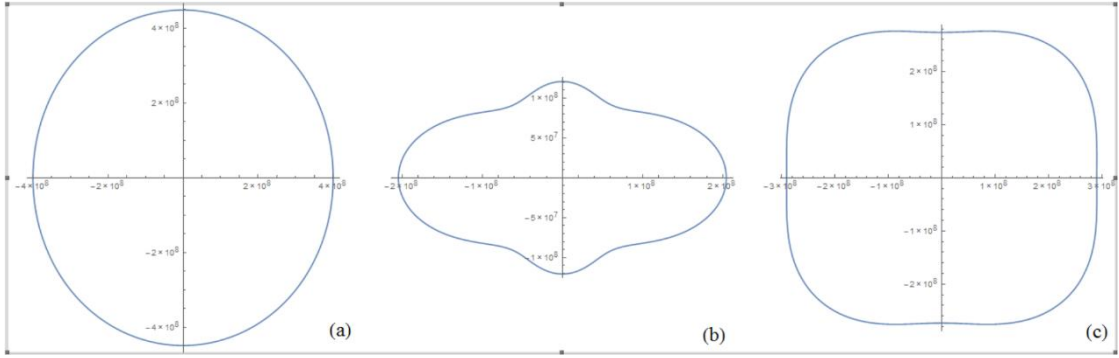

Figure S1. Variation of (a) $h_1$, (b) $h_2$, and (c) $h_3$ with $\alpha$ in polar coordinate.

For convenience of free energy formulation, the $\alpha$−dependence of $h_1, h_2$ and $h_3$ can be avoided by some approximation. Here we define

$$\overline{h_k} = \frac{1}{2}\left(\max_{\alpha \in [0,2\pi)} h_k + \min_{\alpha \in [0,2\pi)} h_k\right), \quad (k = 1,2,3) \tag{S19}$$

and eqs. (3a-c) in the main text are obtained by replacing $h_k$ with $\overline{h_k}$ in eqs. (S17a-c).

## 2. Landau theory for the wrinkling phase transition of graphene

Within the SCH model, the free energy per atom for graphene with spontaneous curvature is formulated in the main text as

$$f_{SCH} = \langle \Phi_{i0}\rangle + \frac{9}{16}\langle c_a\rangle \tilde{\kappa}^2 + \frac{9}{256}(3\langle c_{a2}\rangle - \langle c_a\rangle)\tilde{\kappa}^4 + \frac{s_0(k_BT)^3}{4\pi\hbar^2}\left(\frac{1}{\langle b_1^*\rangle} + \frac{1}{\langle b_2^*\rangle}\right)J_1 + \frac{s_0(k_BT)^2}{8\pi\hbar}\frac{1}{\sqrt{\langle d_3\rangle}}J_2. \tag{9}$$

When all the independent standard deviations $\sqrt{\Lambda_{L1}}, \sqrt{\Lambda_{T1}}$ and $\sqrt{\Lambda_{Z1}}$ are regarded as small quantities compared with the static lattice constant $a_0$, the following expansion of the potential per atom for flat graphene is valid

$$\langle \Phi_{i0}(r,\theta_{ijk})\rangle = \Phi_{i0}(r,\theta_{ijk}) + \mathfrak{D}_L[\Phi_{i0}]\Lambda_{L1} + \mathfrak{D}_T[\Phi_{i0}]\Lambda_{T1} + \mathfrak{D}_Z[\Phi_{i0}]\Lambda_{Z1} + 3\mathfrak{D}_{Z2}[\Phi_{i0}]\Lambda_{Z1}^2, \tag{S20}$$

where $\mathfrak{D}_L[\ ]$, $\mathfrak{D}_T[\ ]$, $\mathfrak{D}_Z[\ ]$ and $\mathfrak{D}_{Z2}[\ ]$ are operators related to the atomic structure and potential of the system. For graphene described by LCBOP II, we have

$$\mathfrak{D}_L[\ ] = \frac{1}{16a^2}\left\{8a^2\frac{\partial^2}{\partial r^2} + 9\left[2\frac{\partial}{\partial \cos\theta_{ijk}} + \frac{\partial^2}{\partial \cos\theta_{ijk}^2} + \frac{\partial^2}{\partial \cos\theta_{ijk}\partial \cos\theta_{ijl}}\right] + 24a\frac{\partial^2}{\partial \cos\theta_{ijk}\partial r}\right\}_{r=a},$$

$$\mathfrak{D}_T[\ ] = \frac{1}{16a^2}\left\{8a\frac{\partial}{\partial r} + 14\frac{\partial}{\partial \cos\theta_{ijk}} + 39\left[\frac{\partial^2}{\partial \cos\theta_{ijk}^2} - \frac{\partial^2}{\partial \cos\theta_{ijk}\partial \cos\theta_{ijl}}\right]\right\}_{r=a},$$ (S21)

$$\mathfrak{D}_Z[\ ] = \frac{1}{2a^2}\left\{a\frac{\partial}{\partial r} + 7\frac{\partial}{\partial \cos\theta_{ijk}}\right\}_{r=a},$$

$$\mathfrak{D}_{Z2}[\ ] = \frac{1}{24a^4}\left\{-3a\frac{\partial}{\partial r} - 81\frac{\partial}{\partial \cos\theta_{ijk}} + a^2\frac{\partial^2}{\partial r^2} + 14a\frac{\partial^2}{\partial \cos\theta_{ijk}\partial r} + \frac{37}{2}\left[\frac{\partial^2}{\partial \cos\theta_{ijk}^2} + \frac{\partial^2}{\partial \cos\theta_{ijk}\partial \cos\theta_{ijl}}\right]\right\}_{r=a}.$$

In eq. (S20), the expansions with $\Lambda_{L1}$ and $\Lambda_{T1}$ are truncated at the first order while the expansion with $\Lambda_{Z1}$ are truncated at the second order. This is because the out-of-plane fluctuation of 2D materials is significantly larger than the in-plane fluctuation. If we further assume that the change of equilibrium lattice constant from its static value $da = a - a_0$ is small, eq. (S20) can be further expanded as

$$\langle \Phi_{i0}(r,\theta_{ijk})\rangle = \{\Phi_{i0}(r,\theta_{ijk}) + \mathfrak{D}_L[\Phi_{i0}]\Lambda_{L1} + \mathfrak{D}_T[\Phi_{i0}]\Lambda_{T1} + \mathfrak{D}_Z[\Phi_{i0}]\Lambda_{Z1} + 3\mathfrak{D}_{Z2}[\Phi_{i0}]\Lambda_{Z1}^2\}_0$$
$$+ \left\{\frac{\partial}{\partial a}[\Phi_{i0}(a,\theta_{ijk}) + \mathfrak{D}_L[\Phi_{i0}]\Lambda_{L1} + \mathfrak{D}_T[\Phi_{i0}]\Lambda_{T1} + \mathfrak{D}_Z[\Phi_{i0}]\Lambda_{Z1} + 3\mathfrak{D}_{Z2}[\Phi_{i0}]\Lambda_{Z1}^2]\right\}_0 da \quad (S22)$$
$$+ \frac{1}{2}\left\{\frac{\partial^2}{\partial a^2}[\Phi_{i0}(a,\theta_{ijk}) + \mathfrak{D}_L[\Phi_{i0}]\Lambda_{L1} + \mathfrak{D}_T[\Phi_{i0}]\Lambda_{T1} + \mathfrak{D}_Z[\Phi_{i0}]\Lambda_{Z1} + 3\mathfrak{D}_{Z2}[\Phi_{i0}]\Lambda_{Z1}^2]\right\}_0 da^2,$$

where $\{Y\}_0$ denotes the value of $Y$ at static lattice constant $a_0$. Eq. (S22) expands $\langle \Phi_{i0}(r,\theta_{ijk})\rangle$ as a truncated polynomial of $da$, where the coefficients are explicitly determined. All the terms with angle brackets can be expanded in the same way as eq. (S22) using this two-step expansion method. If only the lowing order terms are retained, we have for eq. (7) in the main text

$$da = -\left\{\frac{\mathfrak{D}_L[b_3]\Lambda_{L1} + \mathfrak{D}_T[b_3]\Lambda_{T1} + \mathfrak{D}_Z[b_3]\Lambda_{Z1} + 3\mathfrak{D}_{Z2}[b_3]\Lambda_{Z1}^2}{b_{31}}\right\}_0$$
$$-\left\{\frac{h_0 + \mathfrak{D}_L[h_0]\Lambda_{L1} + \mathfrak{D}_T[h_0]\Lambda_{T1} + \mathfrak{D}_Z[h_0]\Lambda_{Z1} + 3\mathfrak{D}_{Z2}[h_0]\Lambda_{Z1}^2}{b_{31}}\right\}_0 \tilde{\kappa}^2, \quad \text{for } \tilde{\kappa}^2 > 0 \quad (S23)$$

where

$$b_{31} = \frac{\partial b_3}{\partial a}. \quad (S24)$$

One should notice eq. (S23) is deduced from a expansion of eq. (7) in the main text to the linear term of $da$ for convience, which gradually loses its accuracy when $|da|$ becomes larger. Using eqs. (S22, S23), it is possible to rewrite eq. (9) as truncated polynomial of $\tilde{\kappa}$:

$$f_{SCH} = (f_{SCH})_0 + A(T)\tilde{\kappa}^2 + B(T)\tilde{\kappa}^4, \quad (13)$$

where $A(T) = A_0 + \gamma_z T \ln\left[\frac{1-e^{-\frac{\Theta_{out}}{T}}}{1-e^{-\frac{\Theta_{min}}{T}}}\right] + \gamma_{z2}\left(T \ln\left[\frac{1-e^{-\frac{\Theta_{out}}{T}}}{1-e^{-\frac{\Theta_{min}}{T}}}\right]\right)^2 + \gamma_2 T^2 + \gamma_3 T^3$,

$B(T) = B_0 + B_T(T)$ and

$$B_0 = \left\{\frac{h_{00}^2}{2b_{31}^2}\frac{\partial^2 \Phi_{i0}}{\partial a^2} + \frac{9}{256}(3c_{a2} - c_a)\right\}_0,$$

$$B_T(T) = \frac{h_{00}^2}{2b_{31}^2}\left\{\frac{\partial^2(\Phi_{i00} - \Phi_{i0})}{\partial a^2}\right\}_0 - \frac{9}{16}\frac{h_{00}}{b_{31}}\left(\left\{\frac{\partial c_{a0}}{\partial a}\right\}_0 + \left\{\frac{\partial^2 c_{a0}}{\partial a^2}\right\}_0 da_0\right) \quad (S25)$$
$$+ \frac{9}{256}\left(\{3c_{a20} - c_{a0}\}_0 + \left\{\frac{\partial(3c_{a20} - c_{a0})}{\partial a}\right\}_0 da_0 - 3c_{a2} + c_a\right.$$
$$\left. + \frac{1}{2}\left\{\frac{\partial^2(3c_{a20} - c_{a0})}{\partial a^2}\right\}_0 da_0^2\right).$$

$$A_0 = \left\{\frac{9}{16}c_a - \sum_{j=L,T,Z}\frac{h_0}{b_{31}}\left[\frac{\partial(\mathfrak{D}_j[\Phi_{i0}]\Lambda_{j10})}{\partial a} - \frac{\partial^2 \Phi_{i0}}{\partial a^2}\frac{\mathfrak{D}_j[b_3]\Lambda_{j10}}{b_{31}}\right]\right\}_0$$
$$- \left\{\frac{h_0}{b_{31}}\left[\frac{\partial(3\mathfrak{D}_{Z2}[\Phi_{i0}]\Lambda_{Z10}^2)}{\partial a} - \frac{\partial^2 \Phi_{i0}}{\partial a^2}\frac{3\mathfrak{D}_{Z2}[b_3]\Lambda_{Z10}^2}{b_{31}}\right]\right\}_0$$
$$+ \frac{9}{16}\left\{da_{00}\frac{\partial c_a}{\partial a} + \sum_{j=L,T,Z}\mathfrak{D}_j[c_a]\Lambda_{j10} + 3\mathfrak{D}_{Z2}[b_3]\Lambda_{Z10}^2\right\}_0,$$

$$\gamma_z = -\left\{\frac{h_0}{b_{31}}\left[\frac{\partial(\mathfrak{D}_Z[\Phi_{i0}]\Lambda_{Z1t})}{\partial a} - \frac{\partial^2 \Phi_{i0}}{\partial a^2}\frac{\mathfrak{D}_Z[b_3]\Lambda_{Z1t}}{b_{31}}\right]\right\}_0$$
$$+ \frac{9}{16}\left\{-\frac{\mathfrak{D}_Z[b_3]\Lambda_{Z1t}}{b_{31}}\frac{\partial c_a}{\partial a} + \mathfrak{D}_Z[c_a]\Lambda_{Z1t}\right\}_0 + \gamma_{zh},$$

$$\gamma_{z2} = -\left\{\frac{h_0}{b_{31}}\left[\frac{\partial(3\mathfrak{D}_{Z2}[\Phi_{i0}]\Lambda_{Z1t}^2)}{\partial a} - \frac{\partial^2 \Phi_{i0}}{\partial a^2}\frac{3\mathfrak{D}_{Z2}[b_3]\Lambda_{Z1t}^2}{b_{31}}\right]\right\}_0 \quad (S26)$$
$$+ \frac{9}{16}\left\{-\frac{3\mathfrak{D}_{Z2}[b_3]\Lambda_{Z1t}^2}{b_{31}}\frac{\partial c_a}{\partial a} + 3\mathfrak{D}_{Z2}[c_a]\Lambda_{Z1t}^2\right\}_0 + \gamma_{z2h},$$

$$\gamma_2 = -\left\{\frac{h_0}{b_{31}}\frac{s_0 k_B^2}{4\pi\hbar}\left(\frac{1}{a_0\sqrt{d_3}} - \frac{d_{31}}{4d_3^{3/2}}\right)J_2\right\}_0,$$

$$\gamma_3 = -\left\{\frac{h_0}{b_{31}}\frac{s_0 k_B^3}{4\pi\hbar^2}\left(\frac{2}{a_0 b_1} + \frac{2}{a_0 b_2} - \frac{b_{11}}{b_1^2} - \frac{b_{21}}{b_2^2}\right)J_1\right\}_0$$
$$- \left\{\sum_{j=L,T}\frac{h_0}{b_{31}}\left[\frac{\partial(\mathfrak{D}_j[\Phi_{i0}]\Lambda_{j1t})}{\partial a} - \frac{\partial^2 \Phi_{i0}}{\partial a^2}\frac{\mathfrak{D}_j[b_3]\Lambda_{j1t}}{b_{31}}\right]\right\}_0$$
$$+ \frac{9}{16}\left\{\sum_{j=L,T}-\frac{\mathfrak{D}_j[b_3]\Lambda_{j1t}}{b_{31}}\frac{\partial c_a}{\partial a} + \mathfrak{D}_j[c_a]\Lambda_{j1t}\right\}_0 + \gamma_{3h}.$$

$$\gamma_{zh} = -\left\{\frac{\mathcal{D}_Z[h_0]\Lambda_{Z1t}}{b_{31}}\left[\frac{\partial(\mathcal{D}_Z[\Phi_{i0}]\Lambda_{Z10})}{\partial a} - \frac{\partial^2\Phi_{i0}}{\partial a^2}\frac{\mathcal{D}_Z[b_3]\Lambda_{Z10}}{b_{31}}\right]\right\}_0,$$

$$\gamma_{z2h} = -\left\{\frac{3\mathcal{D}_{Z2}[h_0]\Lambda_{Z1t}^2}{b_{31}}\left[\frac{\partial(3\mathcal{D}_{Z2}[\Phi_{i0}]\Lambda_{Z10}^2)}{\partial a} - \frac{\partial^2\Phi_{i0}}{\partial a^2}\frac{3\mathcal{D}_{Z2}[b_3]\Lambda_{Z10}^2}{b_{31}}\right]\right\}_0, \quad (S27)$$

$$\gamma_{3h} = -\left\{\sum_{j=L,T}\frac{\mathcal{D}_j[h_0]\Lambda_{j1t}}{b_{31}}\sum_{j=L,T}\left[\frac{\partial(\mathcal{D}_j[\Phi_{i0}]\Lambda_{j10})}{\partial a} - \frac{\partial^2\Phi_{i0}}{\partial a^2}\frac{\mathcal{D}_j[b_3]\Lambda_{j10}}{b_{31}}\right]\right\}_0.$$

In eq. (S25), $h_{00} = h_0 + \sum_{j=L,T,Z}\mathcal{D}_j[h_0]\Lambda_{j1t} + 3\mathcal{D}_{Z2}[h_0]\Lambda_{Z1t}^2$ and $\Phi_{i00}$, $c_{a0}$ and $c_{a20}$ (i.e., terms with an additional subscript 0) have similar expressions with $h_{00}$; $da_0 = -\frac{1}{b_{31}}\{\sum_{j=L,T,Z}\mathcal{D}_j[b_3]\Lambda_{j1} + 3\mathcal{D}_{Z2}[b_3]\Lambda_{Z1}^2\}$. In eq. (S26), $b_{11} = \frac{\partial b_1}{\partial a}$, $b_{21} = \frac{\partial b_2}{\partial a}$, $d_{31} = \frac{\partial d_3}{\partial a}$, and $\Lambda_{L10} = \frac{\hbar a^2}{3m}\sqrt{\frac{\pi}{s_0 b_1^*}}$, $\Lambda_{T10} = \frac{\hbar a^2}{3m}\sqrt{\frac{\pi}{s_0 b_2^*}}$, $\Lambda_{Z10} = \frac{\hbar a^2}{2m\sqrt{d_3}}$ and $\Lambda_{L1t} = \frac{a^2 s_0 (k_B)^3}{8\pi m\hbar^2}\frac{J_3}{(b_1^*)^2}$, $\Lambda_{T1t} = \frac{a^2 s_0 (k_B)^3}{8\pi m\hbar^2}\frac{J_3}{(b_2^*)^2}$, $\Lambda_{Z1t} = \frac{a^2 s_0 k_B}{8\pi m d_3}$. One should keep in mind that the calculation of $\Lambda_{Z1}$, $\Lambda_{L1}$ and $\Lambda_{T1}$ defined in eq. (10) in the main text has to be achieved in the self-consistent way since the terms $\langle b_1^*\rangle$, $\langle b_2^*\rangle$ and $\langle d_3\rangle$ are related to $\Lambda_{Z1}$, $\Lambda_{L1}$ and $\Lambda_{T1}$. In constructing the Landau theory, however, we omitted the angle brackets of $\langle b_1^*\rangle$, $\langle b_2^*\rangle$ and $\langle d_3\rangle$ in eq. (10) in the main text so that $\Lambda_{Z1}$, $\Lambda_{L1}$ and $\Lambda_{T1}$ can be directly calculated as a function of temperature. To see that effect of such a simplification, we plot the variation of $\Lambda_{Z1}/a_0^2$, and $\Lambda_{T1}/a_0^2$ with temperature within the Landau theory and the SCH calculation in Figure S2. We observe an overestimation of $\Lambda_{Z1}$, and $\Lambda_{L1}$ in the Landau theory, which is responsible for the rapid increase of $\tilde{\kappa}$ (red curve) as temperature increases shown in Figure 1a and the rapid decrease of lattice constant (red curve) shown in Figure 2a in the main text predicted by the Landau theory. At high temperature (above 1000K), this overestimation finally leads to a vanishing $B(T)$ for which $\tilde{\kappa}$ and $da$ both approach infinity and hence the Landau theory is no-longer applicable.

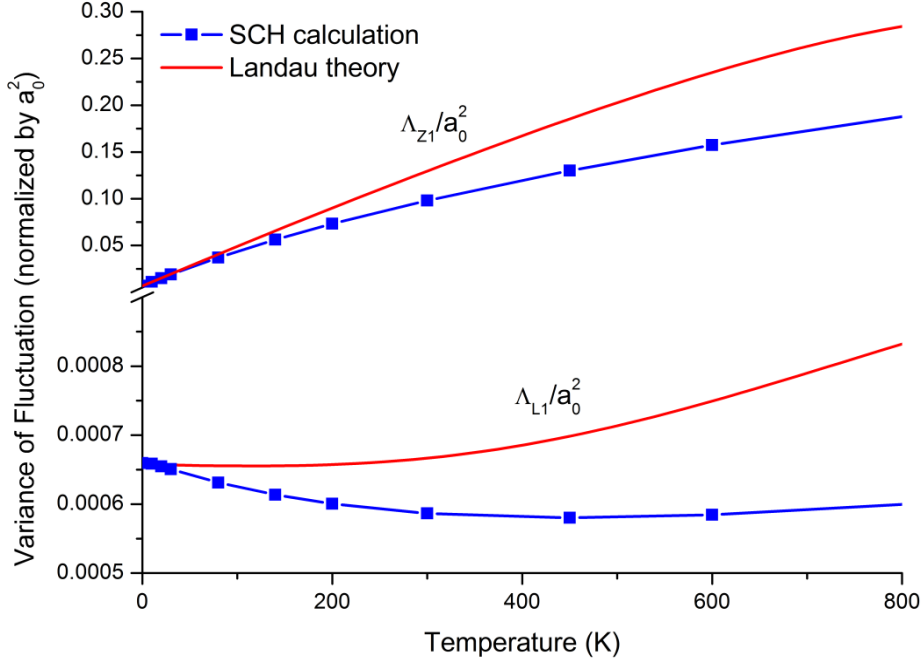

Figure S2. Variation of $\Lambda_{Z1}/a_0^2$, $\Lambda_{L1}/a_0^2$ and $\Lambda_{T1}/a_0^2$ with temperature in the SCH calculation (blue curve with dots) and in the Landau theory (red curve).

To further analyze the origin of the phase transition, $A(T)$ in eq. (13) can also be recasted into the following form

$$A = A_{QH} + A_L + A_T + A_Z, \qquad (S28)$$

where

$$A_{QH} = \left\{\frac{9}{16}c_a - \frac{h_{00}}{b_{31}}\left[\frac{s_0(k_BT)^3}{4\pi\hbar^2}\left(\frac{2}{a_0b_1} + \frac{2}{a_0b_2} - \frac{b_{11}}{b_1^2} - \frac{b_{21}}{b_2^2}\right)J_1 + \frac{s_0(k_BT)^2}{4\pi\hbar}\left(\frac{1}{a_0\sqrt{d_3}} - \frac{d_{31}}{4d_3^{3/2}}\right)J_2\right]\right\}_0,$$

$$A_L = -\left\{\frac{h_{00}}{b_{31}}\left[\frac{\partial(\mathcal{D}_L[\Phi_{i0}]\Lambda_{L1})}{\partial a} - \frac{\partial^2\Phi_{i0}}{\partial a^2}\frac{\mathcal{D}_L[b_3]\Lambda_{L1}}{b_{31}}\right]\right\}_0 + \frac{9}{16}\left\{-\frac{\mathcal{D}_L[b_3]\Lambda_{L1}}{b_{31}}\frac{\partial c_a}{\partial a} + \mathcal{D}_L[c_a]\Lambda_{L1}\right\}_0,$$

$$A_T = -\left\{\frac{h_{00}}{b_{31}}\left[\frac{\partial(\mathcal{D}_T[\Phi_{i0}]\Lambda_{T1})}{\partial a} - \frac{\partial^2\Phi_{i0}}{\partial a^2}\frac{\mathcal{D}_T[b_3]\Lambda_{T1}}{b_{31}}\right]\right\}_0 + \frac{9}{16}\left\{-\frac{\mathcal{D}_T[b_3]\Lambda_{T1}}{b_{31}}\frac{\partial c_a}{\partial a} + \mathcal{D}_T[c_a]\Lambda_{T1}\right\}_0, \qquad (S29)$$

$$A_Z = -\left\{\frac{h_{00}}{b_{31}}\left[\frac{\partial(\mathcal{D}_Z[\Phi_{i0}]\Lambda_{Z1})}{\partial a} - \frac{\partial^2\Phi_{i0}}{\partial a^2}\frac{\mathcal{D}_Z[b_3]\Lambda_{Z1}}{b_{31}} + \frac{\partial(3\mathcal{D}_{Z2}[\Phi_{i0}]\Lambda_{Z1t}^2)}{\partial a} - \frac{\partial^2\Phi_{i0}}{\partial a^2}\frac{3\mathcal{D}_{Z2}[b_3]\Lambda_{Z1t}^2}{b_{31}}\right]\right\}_0$$
$$+ \frac{9}{16}\left\{-\frac{\mathcal{D}_Z[b_3]\Lambda_{Z1t}}{b_{31}}\frac{\partial c_a}{\partial a} + \mathcal{D}_Z[c_a]\Lambda_{Z1t} - \frac{3\mathcal{D}_{Z2}[b_3]\Lambda_{Z1t}^2}{b_{31}}\frac{\partial c_a}{\partial a} + 3\mathcal{D}_{Z2}[c_a]\Lambda_{Z1t}^2\right\}_0.$$

The physical meaning of these coefficients are explained in the main text. Eq. (S23) establishes a relation between $\tilde{\kappa}^2$ and $da$, for which eq. (13) can be reformulated as a function of $da$. Solving $\frac{\partial f_{SCH}}{\partial(da)} = 0$, we have

$$da = \begin{cases} \dfrac{C_{1l}}{C_{2l}}, & T \geq T_c \\ \dfrac{C_{1h}}{C_{2h}}, & T < T_c \end{cases} \qquad (16)$$

where

$$C_{1l} = -\left\{\frac{\partial \Phi_{i00}}{\partial a} + \frac{b_{31}}{h_0}\left(\frac{9}{16}c_a - A_{QH}\right)\right\}_0,$$

$$C_{2l} = \left\{\frac{\partial^2 \Phi_{i00}}{\partial a^2}\right\}_0 + \frac{s_0(k_B T)^3}{4\pi \hbar^2 a_0}\left(\frac{2}{a_0 b_1} + \frac{2}{a_0 b_2} - \frac{4b_{11}}{b_1^2} - \frac{4b_{21}}{b_2^2}\right)J_1 + \frac{s_0(k_B T)^2}{4\pi \hbar a_0}\left(\frac{1}{a_0 \sqrt{d_3}} - \frac{d_{31}}{d_3^{3/2}}\right)J_2,$$

$$C_{1h} = C_{1l} + \left\{\frac{9}{16}\frac{b_{31}}{h_0}\left(c_{a0} - da_0 \frac{\partial c_{a0}}{\partial a}\right) + \frac{9}{128}\frac{b_{31}^2}{h_0^2}\left[da_0(3c_{a20} - c_{a0}) - \frac{da_0^2}{2}\frac{\partial(3c_{a20} - c_{a0})}{\partial a}\right]\right\}_0, \quad (S30)$$

$$C_{2h} = C_{2l} + \left\{\frac{9}{8}\frac{b_{31}}{h_0}\left(-\frac{\partial c_{a0}}{\partial a} + \frac{da_0}{2}\frac{\partial^2 c_{a0}}{\partial a^2}\right) + \frac{9}{128}\left(\frac{b_{31}}{h_0}\right)^2 (3c_{a20} - c_{a0})\right\}_0$$

$$+ \left\{\frac{9}{64}\left(\frac{b_{31}}{h_0}\right)^2\left[-da_0\frac{\partial(3c_{a20} - c_{a0})}{\partial a} + \frac{da_0^2}{4}\frac{\partial^2(3c_{a20} - c_{a0})}{\partial a^2}\right]\right\}_0.$$

S1. Maradudin AA, Montroll EW, Weiss GH, Ipatova I. *Theory of lattice dynamics in the harmonic approximation*, vol. 3. Academic press New York, 1963.

S2. Los JH, Ghiringhelli LM, Meijer EJ, Fasolino A. Improved long-range reactive bond-order potential for carbon. I. Construction. *Physical Review B* 2005, **72**(21).

S3. Stewart GW. *Matrix perturbation theory*, 1990.